\documentclass{elsart1p}

\usepackage{graphics}

\usepackage{amssymb,amsmath}

\def\x{{\mathbf x}}

\begin{document}

\begin{frontmatter}



\title{Semiclassical calculation of decay rates}


\author[usp]{A. Bessa}, 
\author[ufrj]{C.A.A. de Carvalho}, and 
\author[ufrj]{E. S. Fraga}

\address[usp]{Instituto de F\'{\i}sica, Universidade de S\~ao Paulo\\
C.P. 66318, S\~ao Paulo, SP, 05315-970, Brazil}
\address[ufrj]{Instituto de F\'{\i}sica, Universidade Federal do Rio de Janeiro\\
C.P. 68528, Rio de Janeiro, RJ 21941-972, Brazil}

\begin{abstract}
Several relevant aspects of quantum-field processes can be well described by semiclassical methods. In particular, the knowledge of non-trivial classical solutions of the field equations, and the thermal and quantum fluctuations around them, provide non-perturbative information about the theory. In this work, we discuss the calculation of the one-loop effective action from the semiclasssical viewpoint. We intend to use this formalism to obtain an accurate expression for the decay rate of non-static metastable states.

\end{abstract}

\end{frontmatter}

\section{Introduction}
\label{introduction}
A theory for the description of metastable states was formulated by J. Langer long ago\cite{Langer:1967ax,Langer:1969bc}. The formalism and its quantum extensions\cite{Callan:1977pt,Affleck:1980ac} revealed the connection between the decay rate and the free energy of a saddle-point configuration $\phi_s$ of the Euclidean action with a single negative eigenvalue. In general, $\phi_s$ interpolates the local minimum (the false, metastable vaccum) and the global minimum (the true vacuum) of the action. The decay rate is given by
\begin{align}\label{gamma}
\Gamma \;=\;\Omega\,\exp \,-S_E(\phi_{s})\;,
\end{align}
where the pre-factor $\Omega$ is formally written in terms of the determinant of (quantum and thermal) fluctuations around $\phi_s$. In the end, we are led to the problem of calculating the one-loop effective action around $\phi_s$. In this paper, we present a finite-temperature semiclassical procedure to obtain the pre-factor of the decay rate as an alternative to the traditional approach which uses Matsubara sums\cite{Linde:1981zj,Linde:1980tt,Gleiser:1993hf}. Our approach appears more appropriate to be generalized to the case of  non-static saddle-point configurations.

\section{The semiclassical method at finite T}


Semiclassical methods have been successfully applied to quantum statistical mechanics. In this approach, the path-integral expression for the partition function
%
%
is calculated using the steepest descent method. Saddle-points of the action are solutions of the euclidean equations of motion, and configurations in the vicinity of these classical solutions dominate the path integral. The contribution of such configurations can be systematically incorporated, defining a semiclassical series. In the particular case of one-dimensional quantum-mechanical systems, it is possible to generate all the terms of the series using the semiclassical propagator which, in turn, is determined by the classical solution\cite{deCarvalho:1998mv}. Surprisingly, the first term of the semiclassical series can already produce accurate results. As an example, let us consider the single-well quartic potential:
\begin{equation}
V(x)=\frac{1}{2}\,m\omega^2x^2+\frac{1}{4}\,\lambda x^4.
\label{vx}
\end{equation}
Table \ref{T1} exhibts the ground-state energy for different values of the coupling $g=\lambda\hbar/m^2\omega^3$ \cite{deCarvalho:1998mv}. We see that the semiclassical quadratic approximation is in good agreement with numerical techniques that used optimized perturbation theory, even for large values of the coupling. This serves as a motivation for the application of the semiclassical aproximation to finite-temperature quantum field theory.

\begin{table}[t!]
\centering
\begin{tabular}{c|c|c|c}\hline
$g$\; & $E_0$(semiclassical)
& $E_0$(exact) & Error($\%$) \\
\hline
0.4\; & 0.559258 & 0.559146 & 0.02 \\
1.2\; & 0.639765 & 0.637992 & 0.28 \\
2.0\; & 0.701429 & 0.696176 & 0.75 \\
4.0\; & 0.823078 & 0.803771 & 2.40 \\
8.0\; & 1.011928 & 0.951568 & 6.34 \\\hline
\end{tabular}
\caption{\label{T1}Ground state energy of the quartic oscillator in quantum mechanics for different values of the coupling $g$ ($\hbar=m=\omega=1$) [from Ref. \cite{deCarvalho:1998mv}].}
\end{table}

\begin{figure}[t]
\begin{center}
\resizebox*{!}{5cm}{\rotatebox{90}{\includegraphics{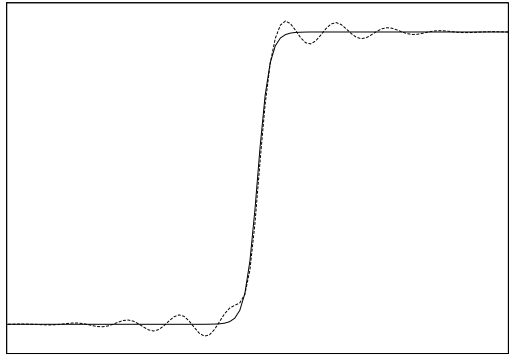}}}
\end{center}
\caption{\label{fig1} Snapshot at $\tau =0$ of profiles contributing to the partition function. The quadratic expansion of the action around the kink (solid line) along the direction of a non-static fluctuation (dashed line) will have non-zero boundary corrections.}
\end{figure}

The path-integral formula for the partition function admits a direct extension to quantum field theories. Indeed, the partition function of a given (scalar) system can be cast in the form:
\begin{equation}\label{eq:Z}
Z=\int[D\varphi(\x)]
\int\limits_{\phi(0,\x)=\phi(\beta,\x)=\varphi(\x)}[D\phi(\tau,\x)]
\;\;\;\; e^{-S_{_{E}}(\phi)}\; ,
\end{equation}
where $S_{_E}(\phi)$ is the Euclidean action of the field:
\begin{equation}
S_{_E}(\phi)=
\int_{0}^{\beta} d\tau d^3\x
\left[\frac{1}{2}\partial_\mu \phi \partial^\mu\phi 
+\frac{1}{2}m^2 \phi^2
+U(\phi)\right]\; .
\end{equation}
We assume that $\phi_s$ is a saddle-point of the action. It obeys the equation of motion
\begin{eqnarray} -\square \phi_s(\tau,\x)+ U'\left(\phi_s(\tau,\x)\right)=0
\; \nonumber\\
 \phi_s(0,\x)=\phi_s(\beta,\x)=\varphi_s(\x)\;,
\label{eq:EOM}
\end{eqnarray}
where we denote by $\square_{_E}\equiv (\partial_\tau^2+{\nabla}^2)$
the Euclidean d'Alembertian operator.  
In principle, $\phi_s$ is a general non-static solution of \eqref{eq:EOM}. Now, we proceed to calculate the contribution to the partition function coming from quadratic fluctuations around $\phi_s$. We write $\phi(\tau,\x) = \phi_s(\tau,\x)+\eta(\tau,\x)$. The only condition on the fluctuation $\eta$ is that $\eta(0,\x) = \eta(\beta,\x)$. In practice, one can restrict the calculation to those configurations with finite action. As an example, one can think of $\phi_s$ as being a kink-like static profile. The finite action condition imposes that $\eta$ goes to zero at spatial infinity. Figure \ref{fig1} ilustrates a typical configuration with finite action in the vicinity of the kink at $\tau =0$. 

A careful expansion of the euclidean action around $\phi_s$ up to quadratic order produces:
\begin{eqnarray}
S_{_E}(\phi)=
 S_{_E}(\phi_s)
+
\delta^{(1)}S_{_E}
+
\delta^{(2)}S_{_E}
+{\cal O}(\eta^3)\; ,
\label{eq:S-expand}
\end{eqnarray}
where 
\begin{eqnarray}\label{eq:1stvariation}
  &&\delta^{(1)}S_{_E}
=\int d^3\x \left[\dot{\phi_s}(x)\eta(x)\right]_{\tau=0}^{\tau=\beta} 
\end{eqnarray}
and
\begin{equation}
\delta^{(2)}S_{_E}\;
=\;
  \frac{1}{2}\int d^3\x
  \left[\eta(\tau,\x)\partial_\tau\eta(\tau,\x)\right]_{\tau=0}^{\tau=\beta}
\;+\;\frac{1}{2}
\int (d^4x)_{_E}\;
\eta(x)\Big[-\square_{_E}+V^{\prime\prime}(\phi_s(x))\Big]\eta(x)\; .
\end{equation}
The boundary terms do not vanish because the boundary value of the fluctuation $\eta$ at $\tau =0,\beta$ is not zero. In other words, there are configurations close to $\phi_s$ whose boundary value is not the same as $\varphi_s(\x) = \phi_s(0,\x)$. In order to integrate (\ref{eq:Z}) over boundary values in the neighborhood of $\varphi_s(\x)$, we can use the techniques of Ref. \cite{Bessa:2007vq}, which incorporate fluctuations of boundary conditions. We write the boundary field as $\varphi(\x) \;=\; \varphi_s(\x) + \xi(\x)$, and expand the action up to quadratic order in $\xi$. To be consistent with the quadratic approximation, we introduce a number of important simplifications which make the problem tractable. It is possible to show that $Z$ around $\phi_s$ is given by the following formula:
\begin{equation}\label{eq:Z2}
Z\;\approx\; e^{-S_{_E}(\phi_s)}
\;(\det G)^{-1/2}\;,
\end{equation}
where
\begin{subequations}\label{eq:propagator}
\begin{eqnarray}
\left[-\square_{_E} + V''\left(\phi_s\right)\right]
G(x;x^\prime)
=\delta^{(4)}(x-x^\prime)\\
G(\tau,\x;0,\x^\prime)=G(\tau,\x;\beta,\x^\prime)=0\;. 
\end{eqnarray}
\end{subequations}
From (\ref{eq:Z2}), we obtain the pre-factor defined in (\ref{gamma}). In special cases, the Green function (\ref{eq:propagator}) can be analytically calculated. For instance, we consider a scalar theory with a quartic potential, and a static kink solution which interpolates between the two equivalent minima:
\begin{equation}
\psi'' + m^2\psi - \frac{\lambda}{4}\psi^3 \;=\;0\;\;\;\rightarrow\;\;\;\psi(x) = \frac{m}{2\sqrt{\lambda}}\tanh\left[\frac{m\,x}{\sqrt{2}}\right ] \;.
\end{equation}
It is possible to show that the Green function we need has the form:
\begin{equation}
G(\tau,\x;\tau ',\x ') = \frac{2}{\beta}\sum_{n=1}^{\infty}\widetilde{G}(\omega_n,\x,\x')\,\sin(\omega_n \tau)\sin(\omega_n \tau ') \;,
\end{equation}
where $\omega_n = \pi n /\beta$. Following \cite{deCarvalho:2001da}, we obtain:
\begin{equation}
\widetilde{G} = \frac{1}{b_n}\left [\rho_+(u)\rho_-(u')\Theta(u'-u) + \rho_+(u')\rho_-(u)\Theta(u-u')  \right ]\;,
\end{equation}
with $\xi = m x/\sqrt{2}$, $b_n = \sqrt{4 + \omega_n^2}$, $u=(1-\tanh \xi)/2$, and 
\begin{equation}
\rho_\pm(u) = \left (\frac{u}{1-u}\right )^{\pm b_n/2}\;p_{\pm}(u)\;,
\end{equation}
where $p_\pm$ are quadratic polynomials. Therefore, we have all the ingredients to calculate the determinant of $G$. Numerical results will be presented in a future publication\cite{future}.

\section{Conclusions}

We presented a systematic procedure to calculate decay rates of metastable states in finite temperature quantum field theory using semiclassical methods. Decay rates are directly related to the one-loop effective action around a saddle point of the Euclidean action.  We illustrated the method in the simple case of a static kink profile. We claim that our approach is particularly useful to deal with non-static saddle-points. 

\vspace{5mm}
\noindent \textbf{Acknowledgment}
\vspace{5mm}

The authors would like to thank the support of CNPq, FAPERJ, FUJB and FAPESP for financial support.



\end{document}